\documentclass[11pt]{article}
\usepackage{epsfig}
\begin{document}

\thispagestyle{empty}

\title{\bf $K^\ast$-couplings for the antidecuplet excitation}
\author{Ya. Azimov$^1$, V. Kuznetsov$^{2,3}$, M. V. Polyakov$^{1,4}$,
I. Strakovsky$^5$\\
\\
$^1$ Petersburg Nuclear Physics Institute,\\ Gatchina, 188300 Russia \\
$^2$Kyungpook National University, \\ Daegu, 702-701, Republic of Korea \\
$^3$ Institute for Nuclear Research,\\ Moscow, 117312 Russia\\
$^4$ Institut f\"ur Theor. Physik -II, Ruhr-Universit\"at,\\
D-44780 Bochum, Germany\\
$^5$ Center for Nuclear Studies, Physics Department, \\ The George
Washington University,\\ Washington, DC 20052 USA}

\date{}
\maketitle

\begin{abstract}
We estimate the coupling of the $K^\ast$ vector meson to the
$N\to\Theta^+$ transition employing unitary symmetry,
vector meson dominance, and results from the GRAAL Collaboration for
$\eta$ photoproduction off the neutron. Our small numerical
value for the coupling constant is consistent with the
non-observation of the $\Theta^+$ in recent CLAS searches
for its photoproduction. We also estimate the $K^\ast$-coupling for
the $N\to\Sigma^\ast$ excitation, with $\Sigma^\ast$ being the
$\Sigma$-like antidecuplet partner of the $\Theta^+\,$-baryon.
\end{abstract}

\section{Introduction}
The experimental status of the exotic baryon $\Theta^+$ is
rather uncertain. Various collaborations have given either positive
results for its observation, or null results, casting doubts on the
existence of the $\Theta^+$ and on the correctness of the positive
claims (see recent experimental reviews~\cite{bur,PDG06}).
The apparently most impressive searches today are the recent CLAS
Collaboration publications~\cite{clas1,clas2,clas3,clas4} with null
results for $\Theta^+$-photoproduction in several final-state
channels.  Nevertheless, investigations are continuing, and several
new (``after-CLAS'') dedicated experiments have given
 both positive~\cite{leps,svd,bar,tr} and
null~\cite{foc,l3,ank,tof,nom} resutls.

The situation is complicated by the fact that there are no data sets,
from independent groups, with exactly overlapping conditions (initial
and/or final states, kinematical regions). Therefore, when
using any null result of one group to reject the positive result of
another, we need to apply some theoretical models/assumptions.
Here we also encounter numerous uncertainties. For
instance, current approaches to the theoretical description of
$\Theta$-production (for example, its photoproduction off the
nucleon~\cite{Tphpr,Close:2004da,Oh:2004wp,Titov:2005kf,Liu:2003hu,
Polyakov:2004aq,Kwee:2005dz}) are mainly based on $K$ and $K^\ast$
meson or reggeon exchanges. The vertex for the $KN\Theta$-coupling
may be considered as known, if one assumes the spin-parity and width
of $\Theta^+$ to be known (the corresponding form factor is still a
problem, of course). Contrary to this, properties of the $K^\ast$
exchange are totally unknown. However, they may be essential,
\textit{e.g.}, for comparisons of $\Theta^+$-photoproduction
off the proton and/or the neutron.

In the present note, we obtain at least rough estimates
of the $K^\ast N\Theta$-coupling. Similar estimates will be provided for
$K^\ast$-vertices contributing to the
excitation $N\to\Sigma^\ast$. To derive the present results, we
relate these vertices to the radiative vertex, extracted
recently~\cite{akps} from preliminary experimental evidence on
$N(1675)$, obtained by the GRAAL Collaboration \cite{Slava}, with
preliminary confirmation from the CB-ELSA-TAPS~\cite{Jaegle} and
LNS~\cite{lns} Collaborations.

\section{Transition vertices}

Let us consider transitions between two baryons $B_1\to B_2$
through radiation of a photon or $K^\ast$. As a first step, we
may assume the flavor symmetry $SU(3)_F$ to be exact. Then, the
corresponding vertices for the vector meson octet are generated in
QCD by the quark vector current $J_{\mu}^{a}$. This is an octet in
$SU(3)_F$, and the superscript $a$ specifies its components. The
$K^\ast$ meson, or any other member of the vector meson octet, is
coupled to a particular combination of the components of the
$J_{\mu}^{a}$.  The electromagnetic current is also related
(proportional) to one of those components. Since it is conserved, and
the $SU(3)_F$ is assumed to be exact, all other components should be
conserved as well.  If both initial and final baryons have $J^P=1/2^+$,
then the transition vertex $V_{\mu}^a(B_2B_1)$ has the general form
\begin{equation}
\label{eq:vert} V_{\mu}^a(B_2B_1)= \langle
B_2|J_{\mu}^{a}|B_1\rangle= \langle
B_2|\left(f_{1}^a(B_1B_2)\gamma_{\mu}+f_{2}^a(B_1B_2)
\frac{i\sigma_{\mu\nu}q^{\nu}}{m_1+m_2}\,+f_{3}^a(B_1B_2)
\frac{q_{\mu}}{m_1+m_2}\right)|B_1\rangle\,,
\end{equation}
where $q=p_2-p_1$ is the momentum transfer, $m_1$ and $m_2$ are the
masses of the corresponding baryons, and the form factors
$f_{i}^a(B_1B_2)$ are invariant functions of $q^2\,$. For a specific
vector meson $M$, we obtain the corresponding vertex $MB_1B_2$ of the
same structure, with coefficients $f_i(MB_1B_2)$ which are specific
combinations of $f_{i}^a(B_1B_2)\,$.

Conservation of the current $J_{\mu}^{a}$ implies that
$$f_{1}^a(B_1B_2)\cdot(m_2-m_1)=- f_{3}^a(B_1B_2)\,\frac{q^2}
{m_1+m_2}\,.$$ In the present note, we are interested in transitions
between baryons of different flavor multiplets and, hence, of
different masses.  For such transitions, $f_{1}^a$ is proportional to
$q^2\,$ and, thus, disappears at $q^2=0\,$, in contrast to $f_{2}^a\,$.
For the case of exact $SU(3)_F$, the last term of the
vertex~(\ref{eq:vert}) does not provide any non-vanishing physically
meaningful contribution (in total analogy with a similar term in the
photon vertex).

Though the situation might change after accounting for the violation
of $SU(3)_F\,$, we advocate the tensor coupling (\textit{i.e.},
the second term of Eq.~(\ref{eq:vert})) to have the leading role at
small $q^2\,$ for non-diagonal transitions. The relative contributions
of other couplings should be suppressed. For the radiative decay
$B_1\to B_2\,\gamma$ (or $B_2\to B_1\gamma$), with the real photon
having $q^2=0$, only the second term in the vertex is physically
meaningful, and \begin{equation} \label{eq:gvert}
f_2(\gamma\,B_1\,B_2)\,|_{q^2=0}=\mu(B_1\to B_2)\cdot(m_1+m_2)\,,
\end{equation} where $\mu(B_1\to B_2)$ is the transition magnetic
moment.

The dominance of the tensor coupling over the vector one in hadronic
non-diagonal transitions (between different flavor multiplets) is
supported by the analysis~\cite{irv} of data for transitions between
$\mathbf{10}$ and $\mathbf{8}$ multiplets. For diagonal transitions
$B\to B$, the conservation of the current $J_{\mu}^{a}$ does not
require $f_{1}^a(B\,B)$ to vanish at $q^2\to 0$, though the ``tensor
dominance'' may still be true as well, at least for some components of
the current. For example, experimental data compilation~\cite{dum}
suggests that for the vertex $\rho NN$ the ratio $f_2/f_1$ is large
($\sim6$, the tensor-over-vector dominance is true), but for the
vertices $\omega NN$ and $\phi NN$ it is small ($< 0.2$, the dominance
is untrue).

Regrettably, many phenomenological calculations of $\Theta$-production
(\textit{e.g.,}~\cite{Tphpr,Close:2004da,
Oh:2004wp,Titov:2005kf,Liu:2003hu}) have used only the vector
coupling for vector-meson exchanges, assuming a non-vanishing
$f_{1}^a|_{q^2\to 0}\,$ and neglecting $f_{2}^a(q^2)\,$. To our
best knowledge,  dominance of the tensor coupling has been
used only in Refs.~\cite{Polyakov:2004aq,Kwee:2005dz}.

\section{$\mathbf {SU(3)_F}$ relations}

Let us consider the tensor couplings for the transitions
$\mathbf{8}\to\overline{\mathbf{10}}$. An octet baryon cannot be
transformed into an antidecuplet member using a unitary singlet meson,
while it can be done through coupling with a unitary octet meson.
Since the product $\mathbf{8\otimes8}$ contains
$\overline{\mathbf{10}}$ only once, all the appearing coupling
constants may be expressed through one of them. The corresponding
relations may be found by means of the Clebsch-Gordan coefficients for
the group $SU(3)$~\cite{deSw}. This method was used in Ref.~\cite{OKL}
to find relations between coupling constants for transitions
$\overline{\mathbf{10}}\to\mathbf8$ with radiation of the octet
pseudoscalar mesons. For our case (vector mesons), these results would
need some modification, because the vector mesons have an essential
singlet-octet mixing, absent for the pseudoscalar mesons. Instead, we
will apply here another (equivalent) way which uses only the
standard (and more widely familiar) $SU(2)$-formalism of the
Clebsch-Gordan coefficients, with the corresponding phase conventions.

Our approach is based on the fact that the group $SU(3)_F$ has
three $SU(2)$-subgroups~\cite{deSw}: the familiar isotopic spin
($I$-spin) group $SU(2)_I$, with the spinor $(u, d)$ and singlet $s$;
the $U$-spin group $SU(2)_U$, with the spinor $(d, s)$ and singlet
$u$; the $V$-spin group $SU(2)_V$, with the spinor $(s, u)$ and
singlet $d$. In the framework of exact $SU(3)_F\,$, the
photon corresponds to the $U$-singlet component of octet.

It is sufficient for our present purpose to use any two of the three
subgroups. Most convenient for us are the isotopic spin and $U$-spin
subgroups (note that all members of any $U$-spin multiplet have the
same electric charge).

With sufficient accuracy for our purpose here, we assume the nonet of
ground state vector mesons to have ideal octet-singlet mixing, such
that $\phi = s\overline s$, while the $\omega$-meson contains light
quarks only, $\omega=(u\overline u+d\overline d)/\sqrt2$. The ground
state baryons are taken, as usual, to be members of an unmixed octet.
For the antidecuplet, we neglect possible mixings, considering them to
be parametrically small~\cite{Diakonov:1997mm}.

In terms of $U$-spin, the proton belongs to the doublet $(p,
\Sigma^+)$, with $U=1/2$, while the proton-like member of the
antidecuplet, $p^\ast$, enters the $U$-spin quartet
$(\Theta^+,p^\ast,\Sigma^{\ast +},\Xi_{3/2}^+)$, with $U=3/2$. The
vector meson nonet contains two independent $U$-singlet combinations,
which cannot change $U$-spin and, thus, cannot transform $p\to p^\ast$.
We may take them as $\rho^0+\omega=\sqrt2\,u\overline u$ and
$\omega+\phi/\sqrt2=(u\overline u+d\overline d+s\overline s)/\sqrt2$
(note that the $U$-singlet component of an octet may change the
isotopic spin, as does the photon).

With the above $U$-singlet combinations, we immediately obtain two
relations \begin{equation} \label{eq:romph}
f_2(\rho^0\,p\,p^\ast)=-f_2(\omega\,p\,p^\ast)=
\frac1{\sqrt{2}}\,f_2(\phi\,p\,p^\ast)\,. \end{equation}
The meson $K^{\ast 0}=d\overline s$ has $U=1$ with $U_3=+1$. It is a
member of a $U$-spin triplet, other members of which are the
combination $(\omega-\rho^0)/2- \phi/\sqrt{2}=(d\overline d-s\overline
s)/\sqrt2\,$, with $U_3=0\,$, and $(-{\overline K}^{\ast 0})\,$, with
$U_3=-1\,$. When applying to this triplet the standard $SU(2)$
Clebsch-Gordan coefficients, which couple doublet ($J=1/2$) and
triplet ($J=1$) into quartet ($J=3/2$), and accounting for relations
(\ref{eq:romph}), we obtain a simple new relation
\begin{equation} \label{eq:kst} f_2(K^{\ast
0}\,p\,\,\Theta^+)=-\sqrt{6}\,f_2(\rho^0\,p\,p^\ast)\,. \end{equation}
It is then easy to use the standard isotopic ($I$-spin) relations
and obtain the neutron couplings \begin{equation} \label{eq:neut}
f_2(K^{\ast +}\,n\,\,\Theta^+)=-f_2(K^{\ast 0}\,p\,\,\Theta^+),~~~
f_2(\rho^0\,n\,n^\ast)=-f_2(\rho^0\,p\,p^\ast),~~~
f_2(\omega\,n\,n^\ast)=f_2(\omega\,p\,p^\ast)\,,
\end{equation}
where $n^\ast$ is the $n$-like member of the antidecuplet. The
proton-neutron relation for the $\phi$-couplings is similar to that for
$\omega$.

In the same manner, one can find $K^\ast$-couplings for transitions
between the nucleon and $\Sigma^\ast$, the $\Sigma$-like members of the
antidecuplet. According to usual isotopic relations
\begin{equation} \label{eq:sig}
f_2({\overline K}^{\ast 0}\,p\,\,\Sigma^{\ast +})=
-f_2(K^{\ast -}\,n\,\,\Sigma^{\ast -})= -\sqrt2\,f_2(K^{\ast -}\,p\,
\,\Sigma^{\ast 0})=\sqrt2\,f_2({\overline K}^{\ast 0}\,n\,\,
\Sigma^{\ast 0})\,,
\end{equation}
while $\,U$-spin relations give
\begin{equation} \label{eq:sigt}
f_2(K^{\ast 0}\,p\,\,\Theta^+)=-\sqrt3\,
f_2({\overline K}^{\ast 0}\,p\,\, \Sigma^{\ast +})\,.  \end{equation}
Evidently, all the couplings may indeed be expressed through one of
them.  We can take, for example, $f_2(\rho^0\,n\,n^\ast)$ and then find
\begin{equation} \label{eq:sigr}
f_2({\overline K}^{\ast 0}\, p\,\,\Sigma^{\ast +})=-\sqrt2\,
f_2(\rho^0\,n\,n^\ast)\,, ~~~~f_2(K^{\ast -}\,p\,\,\Sigma^{\ast
0})=f_2(\rho^0\,n\,n^\ast)\,.  \end{equation}
In turn, the $\rho$-meson coupling may be estimated starting from
the transition magnetic moment.

\section{Vector meson dominance}

A good approximation for non-hard electromagnetic interactions of
hadrons is the hypothesis that ``the entire hadronic electromagnetic
current operator is \textit{identical} with a linear combination of
the known neutral vector--meson fields''~\cite{klz}. This approach
is now known as the vector--meson dominance (VMD)~\cite{sak} (for
a recent brief review see the talk~\cite{sch}).

The simplest form of VMD takes into account only the lightest mesons
$\rho^0,\, \omega,$ and $\phi\,$. Then the transition magnetic moments
may be expressed as
\begin{equation} \label{eq:magmom}
\mu(B_1\to B_2)=\frac1{m_1+m_2}\,\sum_{V=\rho^0,\,\omega,\,\phi}\,
\frac{e}{g_V}\, f_2(V\,B_1\,B_2)\,|_{q^2=0}\,. \end{equation}
In what follows, we will use $f_2(V\,B_1\,B_2)$ only at $q^2=0\,$,
without showing this explicitly every time.

The meson-photon couplings $g_V$ can be easily related to the partial
widths of decays $V\to e^+e^-$:
\begin{equation} \label{eq:gG}
\frac{g_V^{\,\,2}}{4\pi}= \frac{\alpha^2}3\,\frac{m_V}
{\Gamma(V\to e^+e^-)}\,.
\end{equation}
Assuming exact $SU(3)_F\,$, the photon corresponds to the octet
component with $U=0$. The couplings $g_V$ then satisfy the group
relations
\begin{equation} \label{eq:gV}
\frac1{g_{\rho^0}}:\frac1{g_{\omega}}:\frac1{g_{\phi}}=1:
\frac13:(-\frac{\sqrt{2}}3)\,.
\end{equation}
For illustration, let us consider how these relations, together with
other $SU(3)_F$ relations (\ref{eq:romph}) and VMD relation
(\ref{eq:magmom}), result in the cancellation\footnote{In exact
$SU(3)_F$, $\mu(p^\ast\to p)\,$ should vanish, since the $U$-spins
are 3/2 for $p^\ast$, 1/2 for $p$, and 0 for the photon. With the
violation of $SU(3)_F$, this transition moment becomes non-vanishing,
but is still much smaller than $\mu(n^\ast\to n)\,
$~\cite{Polyakov:2003dx}\,.}
of various contributions to $\mu(p^\ast\to p)\,$:  \begin{equation}
\label{eq:ppst} (m_p+m_{p^\ast})\cdot\mu(p^\ast\to
p)=\frac{e}{g_{\rho^0}}\, f_2(\rho^0\,p\,p^\ast)\left(1 -\frac13-
\frac23\right)=0\,.  \end{equation} Relations (\ref{eq:gV}), with an
additional assumption $m_V^{(0)} = m_V^{(8)}$, also predict that
\begin{equation} \label{eq:Gee} \Gamma(\rho^0\to
e^+e^-):\Gamma(\omega\to e^+e^-):\Gamma(\phi\to e^+e^-)=9:1:2\,.
\end{equation}
Experimentally~\cite{PDG06}, these ratios are
$$(7.0~\rm{keV}):(0.6~\rm{keV}):(1.3~\rm{keV})=
11.6:1:2.1\,.$$
Evidently, $SU(3)_F$-violations are here less than $30\%$. Taking
into account the difference of masses for $\rho^0,\,\omega$, and
$\phi\,,$ one can see that the accuracy of ratios Eq.~(\ref{eq:gV})
is also not worse than $30\%$.  Having in mind the current large
uncertainties in properties of the $N^\ast(1675)$, which we assume to
be the $N$-like partner of the $\Theta^+$, we will take the $SU(3)_F$
relations to be exact.

Then, similar to the proton transition moment (\ref{eq:ppst}),
we obtain the neutron transition moment
\begin{equation} \label{eq:nnst}
(m_n+m_{n^\ast})\cdot\mu(n^\ast\to n) =
\frac{e}{g_{\rho^0}}\,f_2(\rho^0\,p\,p^\ast)\left(-1-\frac13-
\frac23\right)=-\,\frac{2e}{g_{\rho^0}}\,f_2(\rho^0\,p\,p^\ast)=
\frac{2e}{g_{\rho^0}}\,f_2(\rho^0\,n\,n^\ast)\,.
\end{equation}

\section{Numerical estimates}

We are now ready to discuss the numerical values of various quantities.
Experimental characteristics of the $\rho^0$~\cite{PDG06} and
Eq.~(\ref{eq:gG}) give $|g_V|\approx5\,$. Previously, in Ref.~\cite{akps},
we extracted $|\mu(n^\ast\to n)|=(0.13-0.37)\,\mu_N\,$, where
$\mu_N=e/(2m_N)$ is the standard nuclear magneton. This gives
\begin{equation} \label{eq:f2}
|f_2(\rho^0\,n\,n^\ast)|=|f_2(\rho^0\,p\,p^\ast)|\approx(0.13-0.37)\,
\,g_{\rho^0}\,\,\frac{m_N+m_{N^\ast}}{4m_N}\approx(0.45-1.28)\,,
\end{equation}
which is essentially smaller than $|f_2(\rho^0\,N\,N)|$, equal to
$12-16$~\cite{dum}.

With the value in Eq.~(\ref{eq:f2}), Eqs.~(\ref{eq:kst}) and
(\ref{eq:neut}) give
\begin{equation} \label{eq:f2kst}
|f_2(K^{\ast
0}\,p\,\Theta^+)|=|f_2(K^{\ast +}\,n\,\Theta^+)| =
\sqrt{6}\,|f_2(\rho^0\,n\,n^\ast)|=(1.10 - 3.14)\,.
\end{equation}
Analogously, from Eqs.~(\ref{eq:sig}) and (\ref{eq:sigr}), we can
find $K^\ast$-couplings for the $\Sigma^\ast$-excitation:
\begin{equation} \label{eq:kstsig}
|f_2(K^{\ast -}\,p\,\,\Sigma^{\ast 0})|=(0.45-1.28)\,,
~~~|f_2({\overline K}^{\ast 0}\,p\,\,\Sigma^{\ast +})|
=(0.64 - 1.81)\,.
\end{equation}
Incidentally, the value of 1.1 for $|f_2(K^{\ast +}\,n\,\Theta^+)|=
|f_2(K^{\ast 0}\,p\,\Theta^+)|$, supported now by Eq.~(\ref{eq:f2kst}),
was used earlier in Ref.~\cite{Kwee:2005dz}\footnote{In
Ref.~\cite{Kwee:2005dz}, the value of $f_2\approx 1.1.$ has been
obtained from the Chiral Quark-Soliton Model of
Ref.~\cite{Diakonov:1997mm}} to estimate the production cross section
of $\Theta^+$ in photoreactions.  With a $\Theta^+$-width of
1~MeV (in accordance to Refs.~\cite{width}) and the above value of
$f_2(K^\ast\,N\,\Theta)$, calculations~\cite{Kwee:2005dz} find small
cross sections $\sigma_{tot}(\gamma p\to{\overline K}^0\Theta^+)<0.22$~nb
and $\sigma_{tot}(\gamma n\to K^- \Theta^+)<1$~nb, which are below the
limits given recently by the CLAS Collaboration~\cite{clas2,clas4}.

Note the difference between proton and neutron targets. Photoproduction
off the proton is nearly independent of the width of $\Theta^+$, due to
absence of the $K^0$-exchange. Therefore, the cross section of $\gamma
p\to{\overline K}^0\Theta^+$ is mainly determined by the $K^{\ast
0}$-exchange, which is proportional to $|f_2(K^{\ast
0}\,p\,\Theta^+)|^2$ (for more details, see Ref.~\cite{Kwee:2005dz}).
On the other side, the neutron photoproduction $\gamma n\to
K^-\Theta^+$ is mainly determined by the charged $K$-exchange, which is
proportional to $\Gamma_{\Theta^+}$.

\begin{figure}[th]
\centerline{
\includegraphics[height=0.6\textwidth, angle=90]{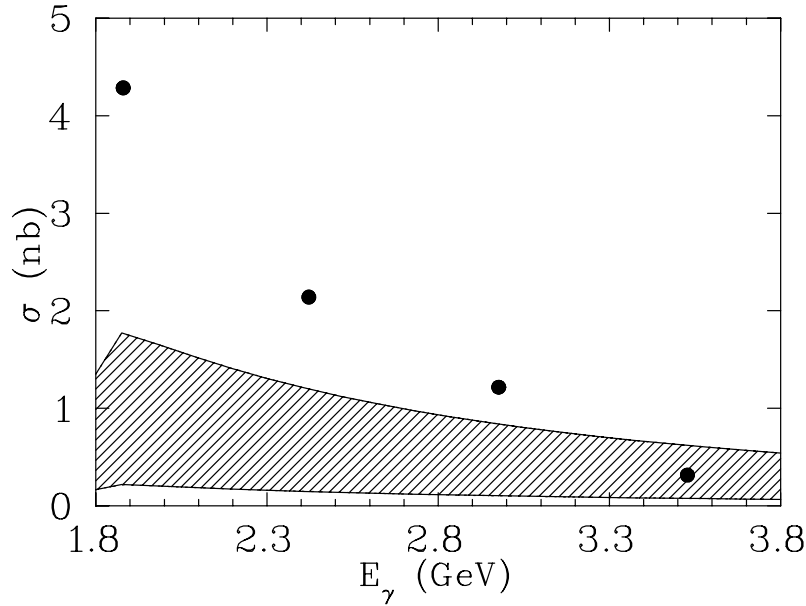}}
\vspace*{0.5cm}
\caption{Shaded band corresponds to the range of $\Theta^+$ production
         cross section in the process $\gamma +p\to{\overline K}^0+
         \Theta^+$ as a function of the photon energy, expected
         according to model~\cite{Kwee:2005dz}. The lower and upper
         edges of the band correspond to the coupling constant
         values $|f_2(K^{\ast 0}\,p\,\Theta^+)|=1.10$ and 3.14,
         the range obtained in the present paper, Eq.~(\ref{eq:f2kst}).
         The filled circles correspond to the experimental upper
         limits for the cross section given by the CLAS
         Collaboration~\protect\cite{clas4}. \label{fig:g1}}
\end{figure}

To clarify the meaning of our numerical values for $f_2\,$, we consider
in more detail the photoproduction
\begin{equation} \label{eq:phthet}
\gamma+p\to \overline K^0+\Theta^+\,, \end{equation}
with $K^*$-exchange as the main contribution. We can compare
experimental limits for this reaction, obtained by the CLAS
Collaboration~\cite{clas4}, and the theoretical calculations
in the model of Ref.~\cite{Kwee:2005dz}. Fig.~\ref{fig:g1} shows the
total cross section of photoproduction (\ref{eq:phthet}) as a function
of the photon energy. Both experimental and theoretical results shown
here assume the same value $m_{\Theta^+}=1540$ MeV. The filled circles
correspond to the experimental upper limits~\protect\cite{clas4} for
particular initial energies. The lower edge of the shaded area
reproduces the lower curve in Fig.7 of Ref.~\cite{Kwee:2005dz}, for
$(J^P)_{\Theta^+}=(1/2)^+$. To obtain the whole shaded area in our
Fig.~\ref{fig:g1}, we multiply this curve by the factor
$|f_2(K^{\ast0}\, p\, \Theta^+)/1.1|^2\,$, with
$f_2(K^{\ast0}\,p\,\Theta^+)$ running the whole above range in
Eq.~(\ref{eq:f2kst}) for the $K^\ast$ coupling constant.

We see that, apart from the higher photon energy region and/or the
upper edge of the interval Eq.~(\ref{eq:f2kst}) for $f_2$, the 
estimate of the cross section is below the limits put by the CLAS
Collaboration. The theoretical results are very rough, of course, 
and contain various uncertainties, both in the approximations used 
and in the phenomenological inputs. Nevertheless, the comparison in
Fig.~\ref{fig:g1} shows that the CLAS analysis procedure of
Ref.~\cite{clas4} might not be sensitive enough to reveal the
$\Theta^+$ (if it exists).

Futhermore, Fig.~\ref{fig:g1} clearly shows that the
$K^\ast$ coupling to the $N\Theta^+$ system should be very small
indeed. Moreover, we expect that corrections
due to a nonzero value of the strange quark mass may further reduce
our present estimates.

\section*{Acknowledgments}

We are grateful to D.~Diakonov for many discussions. We thank
R.~Workman for a critical reading of the manuscript. The work was
partly supported by the Russian State Grant RSGSS-1124.2003.2, by
the Russian-German Collaboration Treaty (RFBR, DFG), by the
COSY-Project J\"ulich, by the Sofja Kowalewskaja Programme of
Alexander von Humboldt Foundation, BMBF, DFG Transregio, by the
U.~S.~Department of Energy Grant DE--FG02--99ER41110, by the
Jefferson Laboratory, and by the Southeastern Universities
Research Association under DOE Contract DE--AC05--84ER40150.


\end{document}